\documentclass[twocolumn,aps,showpacs,amsmath,amssymb,floatfix,pdftex]{revtex4}
\usepackage{graphicx,color}

\begin{document}

\title{Gene Regulation by Riboswitches with and without Negative Feedback Loop}

\author{
  Jong-Chin Lin and
  D. Thirumalai
}

\affiliation{
    Department of Chemistry and Biochemistry,
    Biophysics Program, Institute for Physical Sciences and Technology,
    University of Maryland,
    College Park, MD 20742, U.S.A
}

\date{\today}

\begin{abstract}
Riboswitches, structured elements in the untranslated regions of
messenger RNAs, regulate gene expression by binding specific metabolities.
We introduce a kinetic network model that
describes the functions of riboswitches at the
systems level. Using experimental data for flavin mononucleotide riboswitch 
as a guide
we show that efficient function, implying a large dynamic range without
compromising the requirement to suppress transcription, is determined by a
balance between the transcription speed, the folding and unfolding rates of the
aptamer, and the binding rates of the metabolite.
We also investigated the
effect of negative feedback accounting for binding to metabolites, which
are themselves the products of genes that are being regulated.
For a range of transcription rates negative feedback suppresses gene
expression by nearly 10 fold. Negative feedback speeds the gene expression
response time, and suppresses the change of
steady state protein concentration by half relative
to that without feedback, when there is a modest spike in DNA concentration.
A dynamic phase diagram expressed in terms of transcription
speed, folding rates, and metabolite binding rates predicts different
scenarios in riboswitch-mediated transcription regulation.
\end{abstract}

\maketitle

\section*{Introduction}

Riboswitches are {\it cis}-acting RNA elements located in the untranslated 
region of mRNAs that regulate associated gene expression by sensing and 
binding target cellular metabolites \cite{Winkler2005,Cheah2007,Montange2008}. In bacteria, binding of  
metabolites to the conserved aptamer domain allosterically alters the 
folding patterns of the downstream expression platform, whose conformation 
controls transcription termination or translation initiation 
\cite{Montange2008,Batey2004,Serganov2004}. 
The target metabolites are usually the products or their derivatives of the
downstream gene that riboswitches control. Hence, metabolite binding to
riboswitches serves as a feedback signal to control RNA transcription or
translation initiation.
The feedback through metabolite binding is naturally designed to be 
a fundamental network motif for riboswitches. 
For example, tandem riboswitches respond to multiple metabolites to 
control a single gene with greater regulatory complexity 
\cite{Sudarsan2006,Breaker2008}, while single glmS riboswitch has 
been shown to respond to multiple signals using both negative and positive 
feedback \cite{Watson2011}.
Understanding the various {\it in vivo} riboswitch functions requires a 
theoretical framework that takes into account the interplay between
speed of RNA transcription, folding kinetics of the nascent RNA 
transcript, and kinetics of metabolite binding to the nascent RNA 
transcript, and the role of feedback arising from interactions 
between synthesized metabolities and the transcript.
The effects of speed of RNA transcription and metabolite binding kinetics have been 
examined {\it in vitro} in an insightful study involving the flavin 
mononucleotide (FMN)
riboswitches \cite{Wickiser2005}. They argued that FMN riboswitch is kinetically 
driven implying that the riboswitch does not reach thermodynamic equilibrium 
with FMN before a decision between continued transcription and 
transcription termination needs to be made. 

The regulatory roles played by riboswitches have also inspired design of 
novel RNA-based gene-control elements that respond to small molecules 
\cite{Link2009,Keasling2011}. 
Several models have been proposed to describe how riboswitches  function and meet their  
regulatory demands  \cite{Beisel2009,Chen2009}. 
However, they focused solely on the transcription process without accounting for the feedback effect from the metabolite 
produced by the gene encoding the riboswitch.
Here, we introduce a general kinetic network model
that can be used to describe both {\it in vivo} and {\it in vitro} functions
of riboswitches.
Our coarse-grained kinetic network model, which takes into account the interplay of co-transcriptional 
folding, speed of transcription, and kinetics of metabolite binding, 
also models effects of negative feedback loop 
so that predictions for {\it in vivo} functions of riboswitches can be made.  
As an illustration of the theory, we first consider the dependence of 
metabolite concentration on the regulation of {\it in vitro} transcription 
termination of FMN riboswitches without 
feedback loop, which enables us to obtain the range of 
folding rates and transcription rates that produce results 
consistent with experiments \cite{Wickiser2005}. 
We then include the negative feedback loop in the network to study 
how riboswitches regulate gene expression at the systems level.

\subsection*{General Kinetic Model}
The riboswitch is transcribed from the leader, the non-protein-coding
region, of the associated gene (Fig. 1).
We simplify the multiple complex {\it in vitro} biochemical steps in the function of 
the off riboswitch, involving transcription,
co-transcriptional RNA folding and metabolite binding, to a few key kinetic 
steps (Fig. 1). Without feedback, the first stage is the 
transcription of the aptamer domain 
($B$). The antiterminator sequence is transcribed ($B_{2}$) in the second step. 
At each stage, the aptamer domain of the RNA transcript can either 
co-transcriptionally fold or unfold. Only when the aptamer domain is folded 
($B^{*}$, $B_{2}^{*}$) 
can the RNA transcript bind the metabolite ($M$). At the second stage, when the
aptamer domain is unfolded, the RNA transcript is in an alternative folding
pattern with the formation of antiterminator stem ($B_{2}$). 
The final stage of the transcription occurs when the terminator sequence is 
transcribed ($R_{i}$). If the terminator sequence is transcribed following 
$B_{2}$, the antiterminator structure prevents the formation of terminator
stem and the transcription proceeds till the downstream coding region is 
fully transcribed ($R_{f}$). If the terminator sequence is
transcribed following $B_{2}^{*}$ or $B_{2}^{*}M$, the absence of 
antiterminator allows the terminator to form, which subsequently leads to 
the dissociation of RNA transcript ($B_{2t}^{*}$ and $B_{2t}^{*}M$) from the 
DNA template and terminates the transcription.
The feedback effect involved with translation and metabolite synthesis will be
discussed in later sections. 

\begin{figure*}
\begin{center}
\centerline{\includegraphics*[width=6.5in]{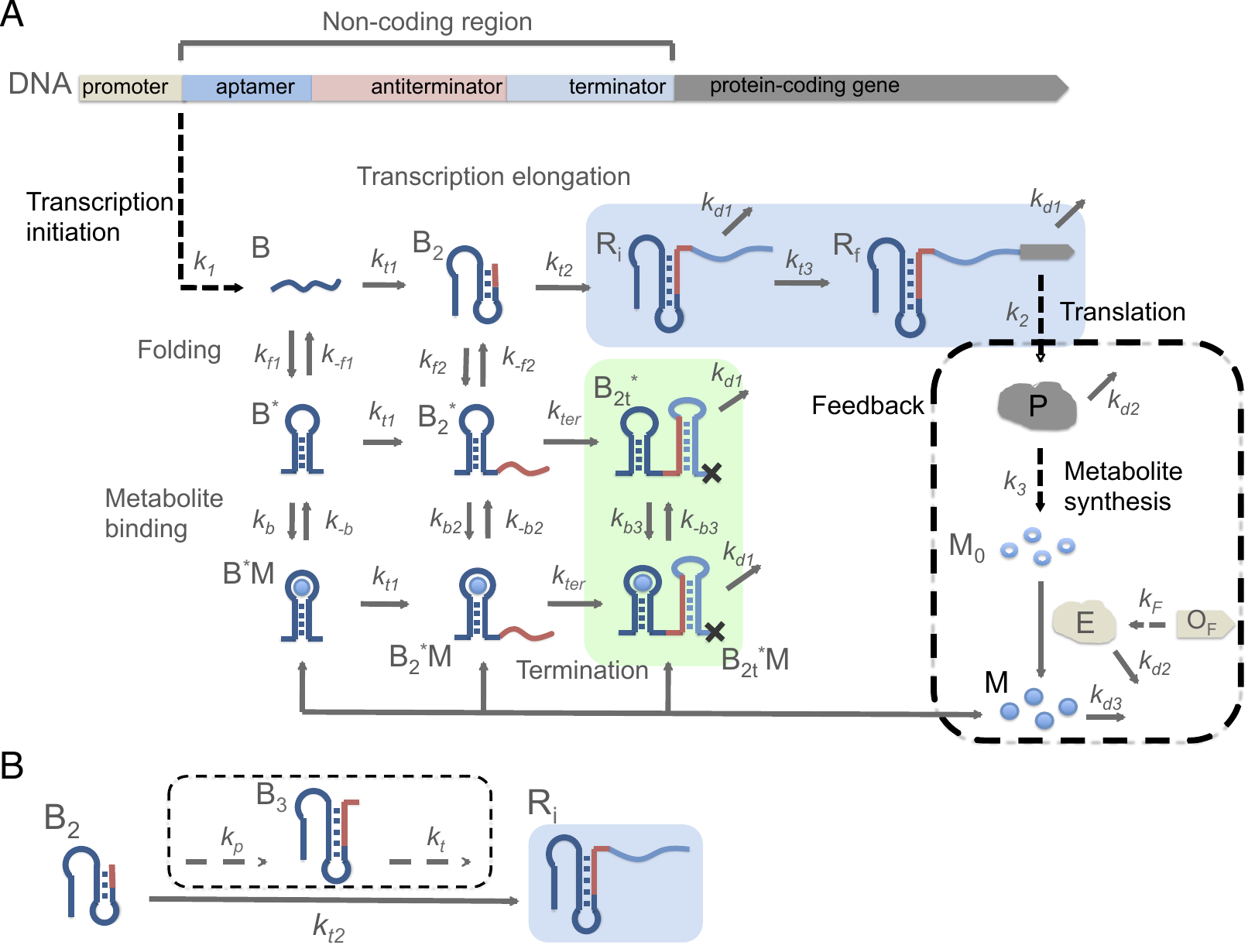}}
\caption{
(A) Kinetic network model for RNA transcription mediated by riboswitches.
The leader, upstream of the protein-coding gene, consists of sequences that
can be transcribed to the aptamer domain ($B$), antiterminator ($B_{2}$) and
terminator region ($R_{i}$) of the riboswitch.
After transcription initiation, elongation, folding of RNA
transcript and metabolite binding are simplified to several key steps.
Starting from the transcript B, where the aptamer sequence is transcribed,
transcription can continue to $B_{2}$ (antiterminator sequence transcribed) at a
transcription rate constant $k_{t1}$. Further elongation through
the terminator sequence with transcription rate constant $k_{t2}$ results in
the synthesis of full RNA without termination. $R_{i}$ is the transcript withthe sequence of the protein-coding region starting to be transcribed, and
eventually grows to $R_{f}$, the full protein-coding region transcribed,
with a rate of $k_{t3}$. Besides transcription elongation, each of the transcript states, $B$ and $B_{2}$, can form
states with aptamer domain folded ($B^{*}$ and $B_{2}^{*}$) with a
folding rate of $k_{f1}$ and $k_{f2}$, respectively.
The aptamer-folded states can bind metabolite ($M$) leading to the
bound states ($B^{*}M$ and $B_{2}^{*}M$) with association rate constant $k_{b}$. The transcripts in state $B_{2}^{*}$
and $B_{2}^{*}M$ can continue elongating till the terminator sequence is
transcribed with their expression platform forming a transcription
terminator stem and dissociate from the DNA template terminating
transcription, with a rate of $k_{ter}$ ($B_{2t}^{*}$ and $B_{2t}^{*}M$).
The fraction of transcription termination, $f_{ter}$, is determined from the
amount of the terminated transcripts (in green block) versus non-terminated
transcripts (in blue block).
In the presence of negative feedback loop (steps included
by the box in dashes) additional biochemical steps have to be included.
In this case after RNA is fully synthesized, it can produce protein $P$ at a
rate $k_{2}$ or get degraded with a rate $k_{d1}$. The fate of $P$ is either
degradation (rate $k_{d2}$) or production of an inactive metabolite $M_{0}$,
which is activated by the enzyme ($E$) encoded by the gene $O_{F}$. The
activated enzyme can bind to the folded aptamer and can abort transcription.
(B) Simplification of the step from $B_{2}$ to $R_{i}$ for FMN riboswitch. In the application to FMN riboswitch, $B_{2}$ represents the transcript out
of the RNA polymerase at the second pause site \cite{Wickiser2005}.
The step $B_{2}\rightarrow R_{i}$ is a simplification of potential multiple chemical process, including pausing and emerging of the
antiterminator sequence ($B_{2}\rightarrow B_{3}$), and transcription to the
terminator sequence ($B_{3}\rightarrow R_{i}$). The rate $k_{t2}$ is
approximated as the pausing rate $k_{p}$, because pausing is likely to be the
rate limiting step in the transcription process.
}\label{Fig1}
\end{center}
\end{figure*}

To assess how the metabolite concentration, $[M]$, regulates transcription
termination, we computed the fraction of terminated transcript, 
$f_{ter}$, given an initial concentration of RNA transcript with aptamer 
sequence transcribed ($B$). Some of the rate constants can be estimated from the 
{\it in vitro} experiments \cite{Wickiser2005} for FMN riboswitches, which we 
use to illustrate the efficacy of the theory. The experimental values of 
the FMN association rate constant $k_{b}$ for the FMN aptamer is 
$\sim 0.1$ $\mu$M$^{-1}$s$^{-1}$, and the dissociation rate constant $k_{-b}$ 
is $\sim 10^{-3}$ s$^{-1}$, giving the equilibrium 
$K_{D}\equiv k_{-b}/k_{b} = 10$ nM. RNA polymerases (RNAP) pause at certain 
transcription sites during transcription. There are two pause sites for the 
FMN riboswitch, one after the aptamer domain
sequence with a lifetime of the paused complex being about 10 s, and the other 
at the end of the antiterminator sequence with a lifetime $\sim 1$ min. 
To approximately account for the pause times in our simplified model, we observe that $B_{2}$ represents
the transcript of the FMN riboswitch with part of the antiterminator      
out of RNAP, when RNAP pauses at the second pause site. Even 
with only part of the antiterminator sequence, the transcript still has high 
probability of forming alternate folding patterns \cite{Wickiser2005}, 
similar to a full antiterminator sequence.
%Therefore, in order to extract the other parameters in our model (Fig. 1)
%by comparison with experiments, 
Hence, we set the effective transcription rates 
$k_{t1} = 0.1$ s$^{-1}$ and 
$k_{t2} = 0.016$ s$^{-1}$, which reflects the pause times for the FMN ribsowtich
(see Fig. 1B for additional explanation for this approximation). 

\subsection*{Extraction of minimal set of parameters from {\it in vitro} transcription experiments} 
To make testable predictions using our model, we need estimates of the co-transcriptional folding and unfolding rates of
the aptamer $B$ as well as $B_{2}$, the aptamer with the antiterminator sequence.
The kinetic model described mathematically in the 
Supplementary Information (SI) can be used to extract parameters that most 
closely fit the measured dependence on  $[M]$ for the FMN riboswitch \cite{Wickiser2005}.  
When the aptamer
sequence is transcribed, the transcript favors the aptamer-folded state, and
when the antiterminator sequence is transcribed, the folding pattern changes in
favor of forming the antiterminator stem with disruption of the aptamer folded
structure \cite{Wickiser2005}. Thus, there are restraints on the folding 
rates, $k_{f1} > k_{-f1}$ and $k_{f2} < k_{-f2}$.
We also assume the same association (dissociation) rate constant 
for metabolite binding to $B^{*}$, $B_{2}^{*}$, and $B_{2t}^{*}$ because
there is little change in the results when the values of $k_{b2}$ and $k_{b3}$
are drastically altered. In addition, because we only allow the folded states of the aptamer to
bind $M$ the effective $K_D$ is a convolution of the folding rates and the 
binding rate. Thus, even though $k_b$ is the same the decrease in the folding 
rate as the transcript length increases effectively decreases $K_D$.
The values of the transitions rates
that reproduce the measured $f_{ter}$ (blue squares in Fig. 2)
are listed in Table I.
The folding rates $k_{f1}$ and $k_{f2}$ are
within an order of magnitude of the theoretical prediction based on, 
$k_f \sim k_0 e^{-\sqrt{N}}$, $k_0 \sim 10^6$ s$^{-1}$, 
where $N$ is the number of nucleotides \cite{Thirum95JPhysI,Hyeon2012BJ}.
Moreover, the rate $k_{f1}$ we obtained is in the same order of 
magnitude of the folding rate of other riboswitch aptamer with similar lengths
observed in experiments \cite{Perdrizet2012}.
Thus, for the purposes of quantitatively describing the {\it in vitro}
experiments we need only two kinetic rates ($k_{-f1}$ and $k_{-f2}$).

\begin{table}[b]
%  \begin{minipage}{4in}
  \begin{center}
  \begin{tabular}[t]{cccccccc} \hline\hline
    && \\[-2.5ex]
    $k_{f1}$\footnotemark[1]
    &\quad
    $k_{-f1}$
    &\quad
    $k_{f2}$
    &\quad
    $k_{-f2}$
    &\quad
    $k_{t1}$\footnotemark[2]
    &\quad
    $k_{t2}$\footnotemark[2]
    &\quad
    $k_{b}$\footnotemark[2]$^{,c}$
    &\quad
    $k_{-b}$\footnotemark[2] \\
    && \\[-2.5ex]
    \hline
    & \\[-1.7ex]
    0.1 &\quad 0.04 &\quad 2.5$\times 10^{-3}$ &\quad 0.04 &\quad 0.1 &\quad 0.016 &\quad 0.1 &\quad $10^{-3}$ \\[-2ex]
    &&\\
    \hline\hline
  \end{tabular}
  \footnotetext[1]{In unit of s$^{-1}$ for all rates except $k_{b}$.}
  \footnotetext[2]{Values from experimental data \cite{Wickiser2005}.}
  \footnotetext[3]{In unit of $\mu$M$^{-1}$s$^{-1}$.}
%  \end{center}
 \caption{Kinetic parameters for model in Figure 1 without feedback.}
\label{table1}
\end{center}
%\end{minipage}
\end{table}

The folding rate of the aptamer is comparable to $k_{t1}$ 
(Fig. 1). The transition rate ($k_{-f2}$) from 
$B_{2}^{*}$ (Fig. 1) to $B_{2}$ is 2-3 times the rate of transcription 
enlongation to the stage where terminator sequence is transcribed ($k_{t2}$). 
Since the results are not sensitive to $k_{-f1}$, if the other rates are
fixed, we choose $k_{-f1}\sim k_{-f2}$ because both
involve unfolding of the folded aptamer structure. With this assumption,
the parameter set that emerge when our model is used to quantitatively describe
(see Fig. 2) the {\it in vitro} kinetic experiments is unique.
In addition, under these conditions, the regulation of transcription
termination works when $[M]$ is in large excess over RNA transcript,
when $\left[M\right]_{0}/\left[B\right]_{0} > 10$.
With $[M] = 1$ $\mu M$ for metabolite concentration and 
$k_{b} = 0.1$ $\mu M^{-1}s^{-1}$, the binding time is of $\sim$10 s, which is 
of the same order of magnitude as the transcription elongation rate and the 
folding rate of the antiterminator stem. Consequently, the metabolite binding 
is unlikely to reach equilibrium before formation of the antiterminator stem. 
Large excess of metabolite, exceeding the equilibrium $K_{D}$ 
($\sim 10$ nM) for FMN binding to $B_{2}^{*}$, over RNA transcript is needed 
for the metabolites to bind the riboswitches with sufficient speed to 
regulate transcription under the conditions explored in experiments 
\cite{Wickiser2005}.

\section*{Results}
\subsection*{Dependence of $f_{ter}$ on $K_2$}
We investigated how $f_{ter}$ depends on  
variations in the transition rates around the parameter set listed in Table I. Figure 2 shows that 
the fraction of terminated transcripts converges to the same value in the limit 
of high metabolite concentration, independent of $K_{2} \equiv k_{-f2}/k_{f2}$,
while keeping the other paramters fixed. At high $[M]$,
$B^{*}$ and $B^{*}_{2}$ are always metabolite bound, which results in 
very low $[B_{2}^{*}]$. Hence, varying $k_{-f2}$ does not affect $f_{ter}$ at 
high $[M]$. In the limit of low $[M]$, $f_{ter}$ decreases as $K_{2}$ 
increases because $k_{-f2}$ exceeds the effective binding rate $k_{b}[M]$
so that $B_{2}$ is preferentially populated. 

The effective metabolite concentration $T_{50}$, at which
\begin{eqnarray}
\frac{f_{ter}(T_{50})-f_{ter}^{L}}{f_{ter}^{H}-f_{ter}^{L}} = 0.5,
\end{eqnarray}
where $f_{ter}^{H} (f_{ter}^{L})$ is the value of $f_{ter}$ at high (low) $[M]$,
does not change much as $K_{2}$ is varied (see the inset in Fig. 2). 
Even when $K_2$ is small $T_{50}/K_D > 1$ implying the concentration
of $[M]$ is in excess of the equilibrium $K_D$ to effect binding to 
$B_{2}^{*}$. 
Since the population of $B_{2}$ is favored as $K_{2}$ increases it follows that
$f_{ter}$ decreases at all concentrations of the metabolite as $K_{2}$ is
changed from a low to a high value (Fig. 2). 
In addition, the transcription rate from 
$B_{2}$ (or $B_{2}^{*}$) to
the next stage where terminator sequence forms ($k_{t2}$) is about one order of
magnitude larger than $k_{f2}$, which means that at low or normal metabolite 
concentration, transitions between $B_{2}$ and $B_{2}^{*}$ states do not reach 
equilibrium before the terminator sequence is transcribed - a result that also follows
from the inequality $T_{50}>K_{D}$.
Finally, the dissociation rate constant is much smaller than $k_{t2}$,
which indicates that once the metabolite is bound, the bound state remains
stable through
transcription termination. Hence, the riboswitch is in the kinetically driven
regime with the parameters used here - a conclusion that was reached in the 
previous study \cite{Wickiser2005}.

\begin{figure}
\begin{center}
\centerline{\includegraphics*[width=3.25in]{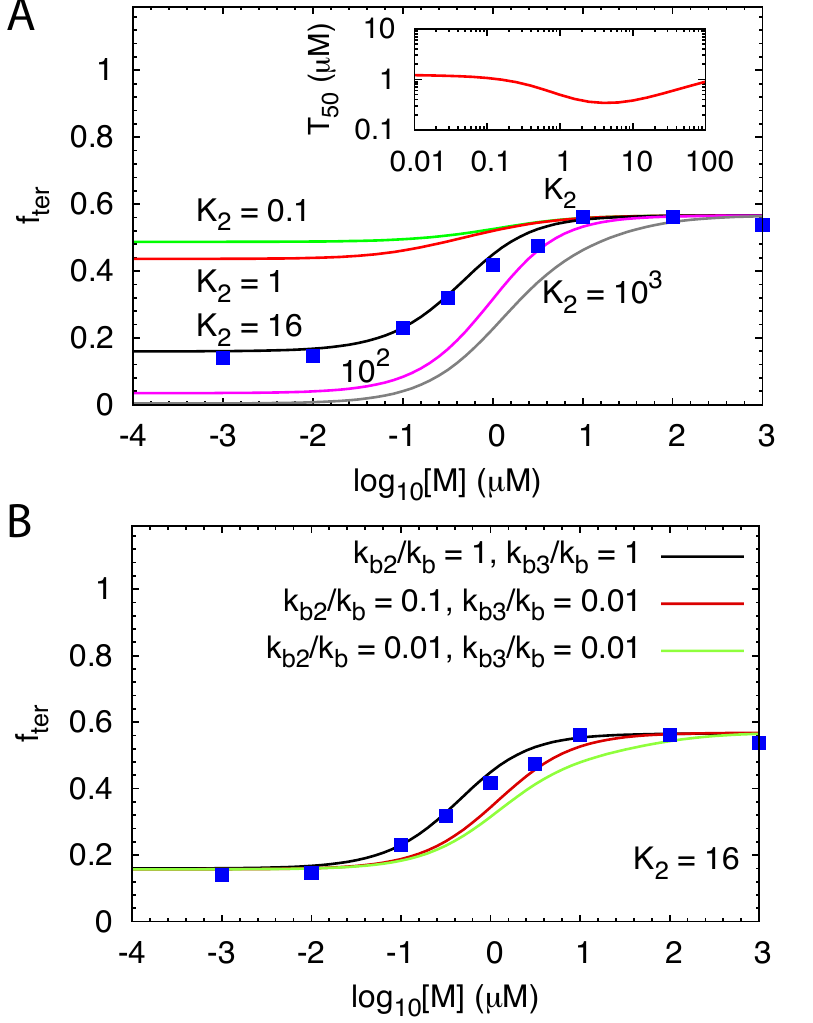}}
\caption{
Dependence of transcription termination on metabolite concentration without feedback.
(A) Fraction of terminated RNA transcripts, $f_{ter}$, as a function of
the logarithm of the metabolite concentration for different values of
$K_{2}\equiv k_{-f2}/k_{f2}$ with $k_{b2}=k_{b3}=k_{b}$. Parameters
that reproduce the {\it in vitro} experimental $f_{ter}$ are listed in Table I.
The inset in (A) shows the half-response
metabolite concentration, $T_{50}$, as a function of $K_{2}$.
(B) Sensitivity of $f_{ter}$ to different
values of $k_{b2}$ and $k_{b3}$. Except for modest changes in $T_{50}$ there is
little change in $f_{ter}$ when the binding rates are drastically altered.
The blue points in (A) and (B) are
data from experiments on FMN riboswitch \cite{Wickiser2005}.
}\label{Fig2}
\end{center}
\end{figure}

\subsection*{Dynamic Range and Thermodynamic Control}
Thermodynamic equilibrium between $B_{2}$ and $B_{2}^{*}$ can be reached only 
if the transcription speed is much slower than the transition rates between 
different folding patterns and the association rate with metabolites (Fig. 1). 
We varied the transcription speed to probe how the riboswitch can be driven 
from kinetic to thermodynamic control, which can be experimentally realized  
by increasing the pausing time, achievable by adding transcription factors, 
such as NusA. The dependence of $f_{ter}$ on $[M]$ at various 
$\gamma_{2} \equiv k_{t2}/k_{f2}$ values shows that, in the limit of low 
metabolite concentration, $f_{ter}$ is roughly equal to the fraction of folded 
aptamer $f_{B_{2}^{*}} \simeq k_{f2}/(k_{f2}+k_{-f2}) \sim 0.06$.
At high metabolite concentrations, almost all riboswitches are metabolite 
bound, and transcription is terminated with high probability (Fig. 3A). 
As $\gamma_{2}$ increases, the system transitons to a kinetically driven 
regime and the probability that the transcription is terminated at high $[M]$ 
decreases, while the fraction of terminated transcript at low $[M]$ increases. 

\begin{figure*}
\begin{center}
\centerline{\includegraphics*[width=6.5in]{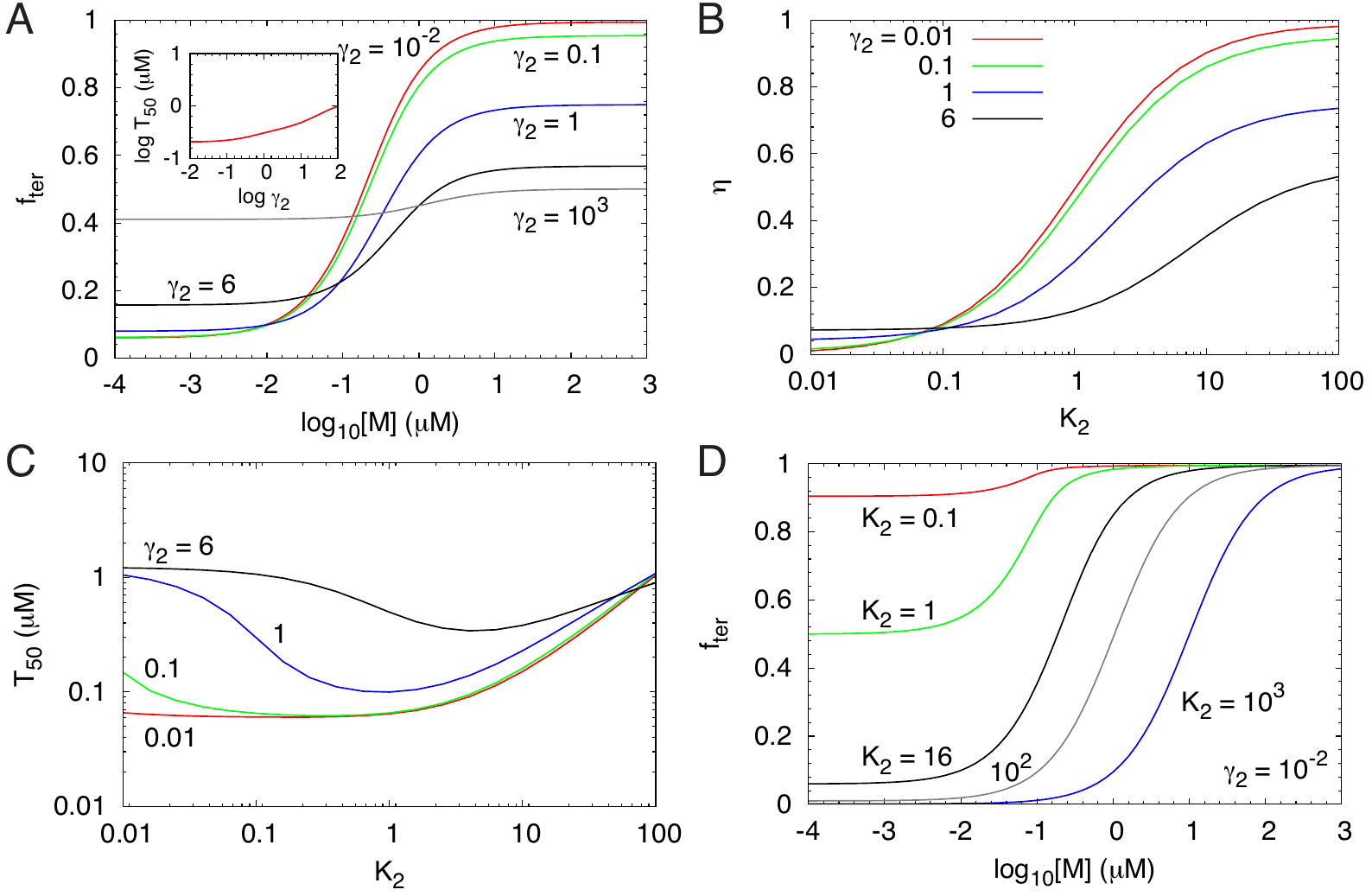}}
\caption{
Speed of transcription and gene expression.
(A) Fraction of terminated RNA transcripts, $f_{ter}$, as a function of
the logarithm of metabolite concentration for different values of
$\gamma_{2}\equiv k_{t2}/k_{f2}$. The parameters that reproduce the
experimental $f_{ter}$ results in $\gamma_{2} = 6$.
The inset shows log($T_{50}$) as a function of log($\gamma_{2}$). 
(B) The dynamic range $\eta \equiv f_{ter}^{H}-f_{ter}^{L}$, where
$f_{ter}^{H}$ ($f_{ter}^{L}$) is the value of $f_{ter}$ at high (low) metabolite
concentration, as a function of $K_{2}$.
(C) Variation of T$_{50}$ as a function of $K_{2}$.
(D) Fraction of terminated RNA transcripts as a function of the logarithm of
metabolite concentration for different values of $K_{2}$ with
$\gamma_{2} = 10^{-2}$. Other parameters used are listed in Table I.
}
\label{Fig3}
\end{center}
\end{figure*} 

Interestingly, at high $[M]$ we find that $f_{ter}$ decreases as $\gamma_{2}$
increases because in this limit the folded $B_{2}^{*}$ has insufficient time
to make a transition to $B_{2}$. As a result the population of $B_{2}$
decreases at high $\gamma_{2}$ resulting in a reduction in $f_{ter}$ (Fig. 3A).
Surprisingly, the exact opposite result is obtained at low $[M]$ as 
$\gamma_{2}$ is varied. At low $[M]$ and small $\gamma_{2}$ the binding rate
$k_{b}[M]$ is small enough that the transition to $B_{2}$ occurs with
high probability
resulting in a decrease in $f_{ter}$. As $\gamma_{2}$ increases the flux from 
$B_{2}^{*}$ to $B_{2}$ decreases, and the
pathway to $B_{2}^{*}$ from $B^{*}$ becomes relevant leading to an increase
in $f_{ter}$ at low $[M]$ (Fig. 3A). Thus, at high $\gamma_{2}$
and low $[M]$ the extent of transcription termination is controlled by
$K_{1}\equiv k_{-f1}/k_{f1}$ and $\gamma_{1}\equiv k_{t1}/k_{f1}$. 
The value of $T_{50}$ increases substantially relative to $K_D$ as $\gamma_2$ 
increases (see the inset in Fig. 3A). Even when $\gamma_2$ is very small 
$T_{50}$ exceeds $K_D$ implying that it is difficult to drive the FMN 
riboswitch to thermodynamic control under the conditions used in experiments 
\cite{Wickiser2005}.

Figure 3B shows the dynamic range  $\eta \equiv f_{ter}^{H}-f_{ter}^{L}$  as
a function of $K_{2}$ for different values
of $\gamma_{2}$. The riboswitch functions with maximal dynamic range
when the system is nearly under thermodynamic control corresponding to small 
$K_{2}$ values.
The range of $T_{50}$ is between 0.1 $\mu M$ ($\approx 100 K_{D}$) and 
1 $\mu M$ ($\approx 1000 K_{D}$)
for $\gamma_{2} > 1$ (Fig. 2C).   When the unfolding rate of the aptamer folded structure 
($B_{2}^{*}$) is 
comparable to the speed of transcription to the terminator sequence, $K_{2} \sim \gamma_{2}$, $T_{50}$ has the smallest value. 
The minimum $T_{50}$ decreases as $\gamma_{2}$ decreases, and 
$T_{50}$ becomes less dependent on $K_{2}$ when $K_{2} < 1$. 
When $\gamma_{2} \ll 1$, as shown
in Fig. 3D, the probability of transcription termination approaches unity in the
limit of high metabolite concentration at all values of $K_{2}$. 
On the other hand, in the limit of low metabolite concentration
($k_{b}[M]$ is small), $f_{ter}$ increases and $T_{50}$ decreases 
as $K_{2}$ decreases. The results in Fig. 3 show that the efficiency of the riboswitch function is determined by a compromise between the need to maximize $\eta$ ($\gamma_2$ should be small) and the ability to terminate transcription ($\gamma_2$ should be large). 

\subsection*{Effect of aptamer folding rates on $f_{ter}$}
In the kinetically driven regime ($\gamma_{2} > 1$), 
the probability of transcription termination
depends on the fraction of the aptamer-formed state before transcription of 
the antiterminator sequence. This fraction can be changed by altering $k_{f1}$,
or equivalently the ratio $K_{1} \equiv k_{-f1}/k_{f1}$, or by varying the 
transcription speed from the aptamer domain sequence to antiterminator 
sequence, $k_{t1}$. When $K_{1} \gg 1$, most of the riboswitches do not form 
stable folded aptamer domain $B^{*}$ (Fig. 1), resulting in very small 
fraction of transcription termination and low response to changes in the 
metabolite concentration. As the transition rate from the unfolded state ($B$) 
to the aptamer-folded state ($B^{*}$) increases relative to the reverse 
transition rate, the fraction of terminated transcripts and dynamic range
increases (Fig. 4A). In the limit of high $[M]$, 
the probability of transcription termination approaches 1 when $K_{1} \ll 1$, 
while in the low $[M]$ limit, almost all the riboswitches are 
aptamer folded but without metabolite bound ($B^{*}$) before transcription to 
antiterminator sequence.  Just as in Fig. 2, $T_{50}$ is not sensitive to changes in $K_1$ (see inset in Fig. 4A).

\begin{figure}
\begin{center}
\centerline{\includegraphics*[width=3.25in]{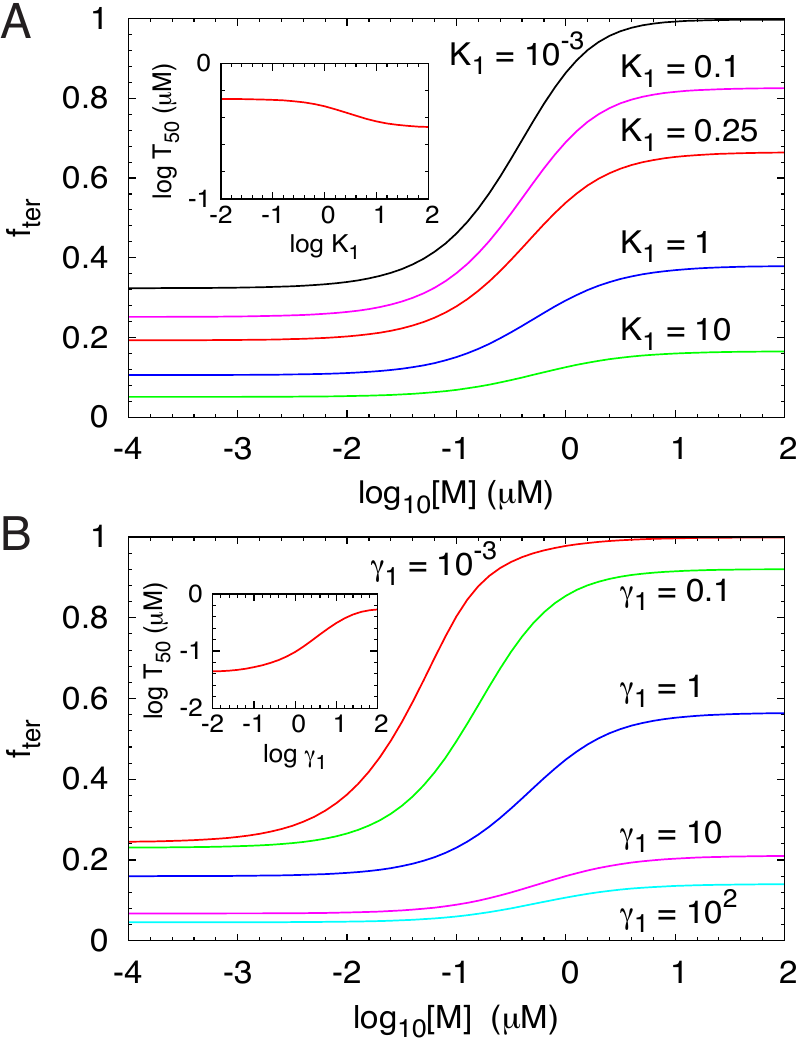}}
\caption{
Aptamer folding rates and $f_{ter}$.
(A) Fraction of terminated RNA transcripts as a function of log($[M]$) for
different values of $K_{1}\equiv k_{-f1}/k_{f1}$ using the parameters
 listed in Table I except for $k_{-f1}$.
(B) Fraction of terminated RNA transcripts as a function of log($[M]$) for different values of $\gamma_{1}\equiv k_{t1}/k_{f1}$.  Inset shows the dependence of $T_{50}$ on $\gamma_1$.
}\label{Fig4}
\end{center}
\end{figure}

Figure 4B shows the concentration dependence of the fraction of terminated
transcripts for different transcription rates to the antiterminator sequence, 
or the ratio $\gamma_{1} \equiv k_{t1}/k_{f1}$. When the transcription rate is 
much faster than the folding rate, the riboswitch does not have enough time to
form the aptamer-folded structure, which results in low fraction of 
terminated transcripts
and low response to metabolite concentration change. When transcription rate 
is much slower than folding rate, the folded and unfolded states of the 
aptamer are able to reach equilibrium before transcription to 
antiterminator sequence. In the high metabolite concentration limit, the 
riboswitch is always metabolite bound resulting in transcription termination.
At low $[M]$, the riboswitch does not bind the metabolite.
The fraction of aptamer formed state without bound metabolite is 
$k_{f1}/(k_{f1}+k_{-f1}) \approx f_{ter}$ before transcription to 
the antiterminator sequence. 
There is substantial variation in $T_{50}$ as 
$\gamma_1$ changes as the inset in Fig. 4B shows. Thus, besides the speed of 
transcription and the binding rates,
 the folding rates of the aptamers have 
considerable influence on $f_{ter}$ (compare insets in Fig. 3A and Fig. 4B).

\begin{table*} 
  \begin{minipage}{5in}
%  \begin{center}
%  \begin{threeparttable}
  \begin{ruledtabular}
  \begin{tabular}[t]{cccccccc}
    && \\[-2.5ex]
    $k_{t3}$\footnotemark[1] 
    &\quad 
    $k_{1}$
    &\quad
    $k_{2}$\footnotemark[2]
    &\quad
    $k_{3}$\footnotemark[3]
    &\quad
    $k_{d1}$\footnotemark[4]
    &\quad
    $k_{d2}$\footnotemark[5]
    &\quad 
    $k_{d3}$\footnotemark[6] 
    &\quad
    $\mu$ \\
    && \\[-2.5ex]
    \hline
    & \\[-1.7ex]
    0.01&\quad 0.016 &\quad 0.3 &\quad 0.064 &\quad $2.3\times 10^{-3}$&\quad $2.7\times 10^{-4}$ &\quad $4.5\times 10^{-3}$&\quad $5\times 10^{-4}$ \\[-2ex]     % 
    &&\\
     
  \end{tabular}
  \end{ruledtabular}
%  \begin{tablenotes}
  \footnotetext[1]{In unit of s$^{-1}$ for all rates.}
  \footnotetext[2]{Ref.\cite{McAdams1997}}
  \footnotetext[3]{Ref.\cite{Schramek2001}}
  \footnotetext[4]{Ref.\cite{Bernstein2002}}
  \footnotetext[5]{Ref.\cite{Belle2006}}
  \footnotetext[6]{See SI.}
%  \end{tablenotes}
%  \end{threeparttable}
%  \end{center}
  \caption{Additional kinetic parameters for model in Figure 1 with negative feedback.}
\label{table2}
%\end{center}
\end{minipage}
\end{table*}

\subsection*{Transcription with negative feedback loop}

Most riboswitches regulate gene expression of the downstream platform that
encodes for proteins involved in the production of the specific metabolite 
that itself binds to the riboswitch (Fig. 1). 
Therefore, sensing and binding of its own metabolite by the riboswitch acts as 
a feedback to control gene expression. 
For riboswitches that suppress gene expression by binding to 
metabolites with high selectivity (for example,
guanine riboswitches or FMN riboswitches) such feedback loop is an example of
negative autoregulation, which has been 
widely studied in gene regulation networks associated with transcription factors
\cite{Alon2007}.  We include the role negative feedback plays 
in regulating transcription termination by generalizing the 
{\it in vitro} kinetic model considered in the previous section.
Our minimal model, illustrated in Fig. 1 and described in detail in the SI,
provides a framework for interpreting future {\it in vivo} experiments.

We consider transcription and translation in a cell and take into
account RNA degradation and cell expansion. 
The transcription process is similar to that 
described without feedback loop, except now we include the 
effect of cell expansion and RNA degradation. 
We assume that the cell grows 
at a rate of $\mu = 5 \times 10^{-4}$ s$^{-1}$, resulting in a typical doubling
time of ln$2/\mu \sim 20$ minutes for an {\it E. coli} cell,  
and that the degradation rate of the fully 
transcribed RNA or terminated RNA transcript is $k_{d1}$.
The values of $k_{d1}$, and other parameters in the feedback loop
are  in Table II.
The fully transcribed RNA serves as a template for the translation of  
the protein ($P$) that synthesizes metabolite $M_{0}$, which is then converted
to an active form $M$ by the enzyme $E$ encoded by the gene $O_{F}$ 
(Fig. 1). 
The species $M$, with a degradation rate of $k_{d3}$, is the target metabolite 
that binds to the riboswitch. 

In the case of FMN riboswitches, $M_{0}$ represents riboflavin, the eventual
product encoded by the gene ribD ($O_{R}$), that 
is subsequently converted to FMN by flavokinase ($E$), synthesized by
the gene ribC ($O_{F}$). The degradation rate $k_{d3}$ takes into account
the effect of conversion from FMN to FAD (flavin adenine dinucleotide) by FAD 
synthetase {\it in vivo} (see SI for more details). However, 
we neglect the potential binding of FAD to the riboswitch 
because there is a 60 fold difference (or potentially even 
larger factor in the absence of FMN) in the binding of FMN and FAD to the FMN 
riboswitch \cite{Winkler02PNAS}.
In the model with negative feedback the extent of regulation by riboswitches
is expressed in terms of the production of the protein $P$ (Fig. 1).

We assume that the activated level of the operon, $O_{R}$, for transcription 
initiation of the riboswitch is a constant, and set 
$\left[DNA\right] = \left[O_{R}\right] = 2.5$ nM, which is equivalent to one 
DNA molecule in an {\it E. coli} cell, and assume that the aptamer 
sequence, $B$, is produced with an effective rate constant 
$k_{1} = 0.016$ s$^{1}$, taking into account the transcription initiation rate 
($\sim 50$ s between initiation events; \cite{Larson2011,Iyer1996}) and the 
typical transcription speed, $\sim 10 - 35$ nucleotides per 
second \cite{Uptain1997}.
With these parameters fixed, which are
used for illustrative purposes only, we can study the effect of feedback by 
varying the effective rate $k_{F}$, at which $E$ is produced from 
O$_{F}$. We set $[O_{F}] = [O_{R}]$. If $E$ is produced at a rate 
similar to that of protein $P$ without feedback, then $k_{F} \sim 1$ s$^{-1}$, 
which we set as a reference rate $k_{F0}$ (see SI for details).
At this rate, the steady state level of enzyme E is $\sim 10^{3}$ copies per
cell.
The variation of rate $k_{F}$ can result from delays or speed up in the 
process of the transcription from $O_{F}$ or translation of E, or
even deficiency in $O_{F}$.

\begin{figure}
\begin{center}
\centerline{\includegraphics*[width=3.25in]{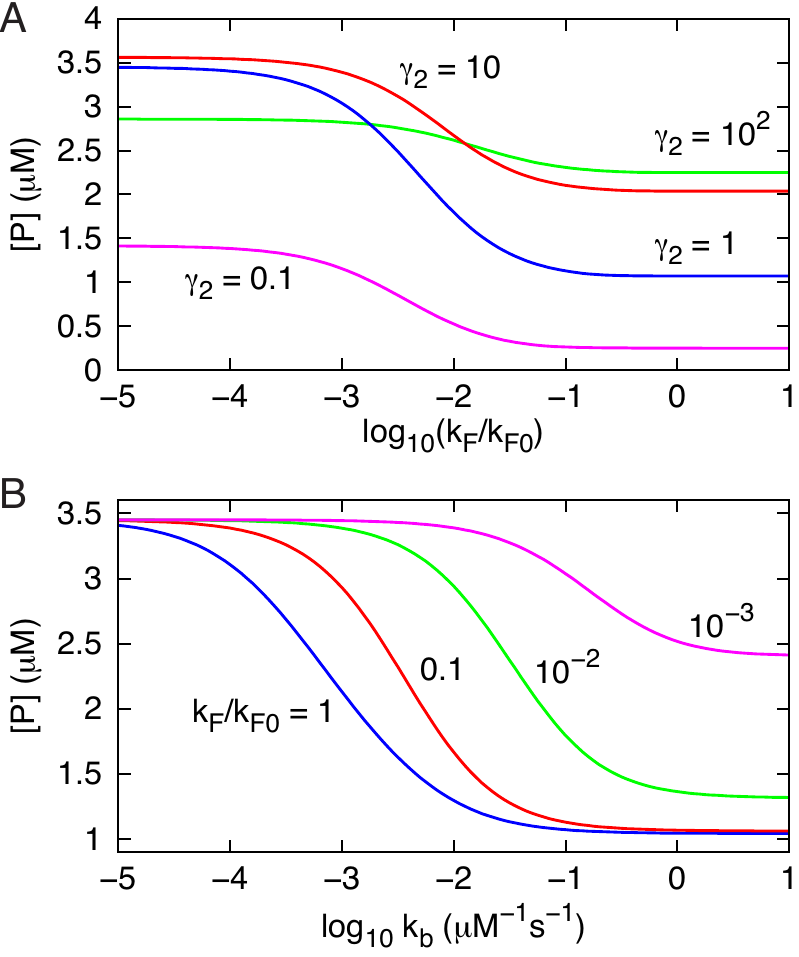}}
\caption{
Effect of negative feedback.
(A) Protein (P in Fig. 1) concentration at different $\gamma_{2}$
values as a function of the logarithm
of $k_{F}$, the production rate constant of enzyme $E$
that produces
the metabolite $M$, relative to $k_{F0}=1$ s$^{-1}$.
The parameters are given in Table I and II except $k_{t2}$.
(B) The extent of regulation expressed in terms of protein level as a
function of the logarithm of association rate constant $k_{b}$ for metabolite
binding. The parameter are listed in Table I and II except $k_{b}$ and $k_{-b}$.
$K_{D}$ is fixed at 10 nM.
}\label{Fig5}
\end{center}
\end{figure}

\subsection*{Dependence of $[P]$ on enzyme production and metabolite binding rates}

We assess the extent of regulation due to feedback by the changes in the
protein level expression, $[P]$, as the parameters in the network are varied.
The results in Fig. 5A show that when $k_{F}/k_{F0}$ is low, very few
active metabolites are formed to suppress protein expression. 
Consequently, the expression level of protein does not change at low
$k_{F}/k_{F0}$ at all values of $\gamma_{2}$ (Fig. 5A).
This finding explains the observation that deficiency in {\it ribC} 
($O_{F}$ in Fig. 1), the gene that encodes for flavokinase (E in Fig. 1), 
causes accumulation of riboflavin ($M_{0}$ in Fig. 1) without converting it to 
FMN, and thus cannot suppress the synthesis of riboflavin \cite{Gusarov1997,Mack1998}. 
When $k_{F}$ increases to about $10^{-3}k_{F0}$, the 
expression of the protein starts to be suppressed. 
There is a substantial suppression of protein concentration (Fig. 5A) at all 
$\gamma_{2}$ values when $k_{F}/k_{F0} > 10^{-2}$. 
As $k_{F}$ exceeds $0.1k_{F0}$, the suppression begins to 
saturate (Fig. 5A) because most of $M_{0}$ produced are converted 
to $M$, and the $[P]$ level is essentially determined by the production of 
$[M]_{0}$, which is independent of $k_{F}$.
Thus, at all $\gamma_{2}$ values, $[P]$ level
varies between two steady state values as $k_{F}$ is changed.

In contrast to the results in Fig. 5A, the extent of regulation ($[P]$ levels) 
varies greatly with binding rate constant of metabolites while keeping
$K_{D}$ fixed at 10 nM (Fig. 5B).
Changes in $[P]$ as $k_{b}$ is varied, which affect synthesis of $M_{0}$
(Fig. 1), depend on $k_{F}/k_{F0}$.
When $k_{F} \geq k_{F0}$, most of the aptamer folded structures are metabolite
bound at experimental value of $k_{b} = 0.1$ $\mu$M$^{-1}$s$^{-1}$, and hence 
the level of protein expression is suppressed. Therefore, with $\sim 10^{3}$ 
copies of enzyme $E$ in a cell, the production of $[P]$ decreases 
substantially even if the binding rate constant is small. Binding occurs because the
 concentration of active metabolites ($\sim 25$ $\mu$M) is far larger than 
RNA transcripts ($\sim 10$ nM),
resulting in a high effective binding rate $k_{b}[M]$.
The ability to suppress protein production decreases if the binding rate
constant is smaller than the experimental value by more than one order of
magnitude, or when the value of $k_{F}$ decreases. Not surprisingly, 
when $k_{F}$ is very low, the dependence of binding rate on expression of $P$ 
decreases. As a consequence the changes in $P$ production decreases greatly as 
$k_{F}/k_{F0}$ decreases (Fig. 5B). Thus, only over a small range of $k_b$ and
$k_{F}/k_{F0}$ does the riboswitch function with sufficient dynamic range.

\begin{figure}
\begin{center}\centerline{\includegraphics*[width=3.25in]{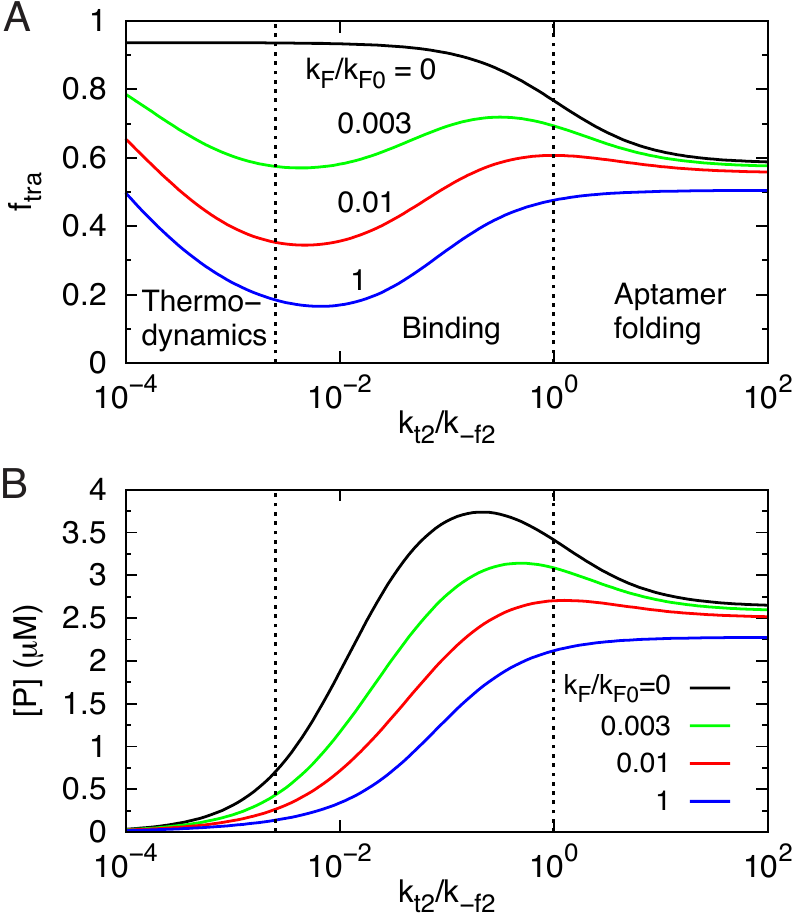}}
\caption{
Role of $k_{t2}$ on negative feedback.
(A) The fraction of fully transcribed RNA, $f_{tra} = 1-f_{ter}$, as a function
of $k_{t2}/k_{-f2}$. The numbers give values of $k_{F}/k_{F0}$
with $k_{F0} = 1$ s$^{-1}$. The dependence of $f_{tra}$ on $k_{t2}/k_{-f2}$
shows three regimes: thermodynamic controlled regime for low $k_{t2}$
($k_{t2}/k_{-f2} < 0.025$, or $k_{t2} < 0.1k_{-b}$), aptamer folding
dominated regime for high $k_{t2}$ ($k_{t2} > k_{-f2}$), and intermediate
regime with significant metabolite binding.
(B) The $P$ concentration as a function of $k_{t2}/k_{-f2}$ at different
$k_{F}/k_{F0}$ values (as described in (A)).
}\label{Fig6}
\end{center}
\end{figure}

\subsection*{Interplay between co-transcriptional folding and transcription speed}
The dependence of transcription rate on the extent of regulation, shown in 
Fig. 6, exhibits three distinct functional modality depending on 
the value of $k_{t2}$ relative to $k_{-f2}$. 
First, consider the case with $k_{F} = 0$ (black line in Fig. 6A). 
If $k_{t2} \gg k_{-f2}$, the fraction of fully transcribed RNA (Fig. 6A), 
$f_{tra} = 1 - f_{ter}$, becomes
\begin{eqnarray}
f_{tra} \equiv \frac{\left[RNA\right]}{\left[RNA\right]_{0}} \simeq \frac{K_{1}\left(1+\frac{k_{t1}+\mu}{k_{-f1}}\right)}{1+K_{1}\left(1+\frac{k_{t1}+\mu}{k_{-f1}}\right)},
\end{eqnarray}
where $\left[RNA\right]_{0}$ is the sum of concentration of the fully 
transcribed and terminated RNA transcripts, and $K_{1}\equiv k_{-f1}/k_{f1}$. 
In this limit, $f_{tra}$ depends predominantly on the folding transition rate
before the antiterminator sequence ($B_{2}$) is transcribed (Fig. 1). 
Hence, $f_{tra}$ is a function of $k_{t1}$, $k_{-f1}$, and the rate of cell 
expansion.
When $k_{t2}/k_{-f2}$ decreases, co-transcriptional folding results in the 
formation of the antiterminator stem, which prevents transcription termination. 
If $k_{t2} \ll k_{-f2}$ and $\mu \ll k_{-f2}$, then
\begin{eqnarray}
f_{tra} &\simeq& \frac{K_{2}}{K_{2}+1} = \frac{k_{-f2}}{k_{f2}+k_{-f2}},
\end{eqnarray}
where $K_{2}\equiv k_{-f2}/k_{f2}$.
In this limit, $B_{2}$ and $B_{2}^{*}$ (Fig. 1) are in equilibrium.
These results are the same as those in the limit of low metabolite 
concentration without the feedback loop (Fig. 3A). 

For finite $k_{F}$, when $k_{t2}$ is fast relative to $k_{-f2}$
the expression level of protein $P$ is nearly
independent of $k_{t2}$ (Fig. 6B).
In this regime transcription termination and hence the extent of 
completed transcription is determined by the folding rates of the aptamer, 
which is not greatly affected by negative feedback. When the transcription
speed decreases, the expression of protein $P$ increases for small $k_{F}$ 
(Fig. 6B).
The expression level of $P$ reaches a peak 
when $k_{t2}/k_{-f2} \sim 0.1-1$, and it  disappears when
$k_{F}/k_{F0} \geq 1$ 
because the metabolite binding becomes significant
enough to stabilize the folded aptamer structure and offset the effect of 
formation of the antiterminator stem. This is illustrated in Fig. 6A, which 
shows that $f_{tra}$ decreases significantly when $k_{t2}$ goes below 
$0.1k_{-f2}$ and reaches a minimum at $k_{t2}/k_{-f2} \sim 0.01$.
The dependence of $[P]$ on $k_{t2}/k_{-f2}$ is maximal when 
$k_{t2}/k_{-f2} \sim 0.01-0.1$.
When the transcription rate decreases further 
($k_{t2} \leq 0.1k_{-b}$, or $k_{t2}/k_{-f2} \leq 2.5\times 10^{-3}$ as shown
by the left dashed line in Fig. 6A), the fraction of fully 
transcribed RNA increases sharply because the transcription rate is slow 
enough for the dissociation of metabolites from riboswitches to occur 
significantly. The system can establish thermodynamic equilibrium, which 
increases the favorability of the formation of antiterminator stem when
$k_{t2}$ decreases resulting in decreasing effective binding rate $k_{b}[M]$. 
However, the overall expression level of protein $P$ becomes very low because 
slow transcription results in a decrease in $P$ production.
In addition, there is also significant probability of RNA degradation,
which also results in a decrease in $P$ expression. Therefore, 
the extent of regulation due to the negative feedback of metabolite binding 
has a maximal effect when $k_{t2}/k_{-f2} \sim 0.01$ where the protein 
expression is suppressed by metabolite binding by as large as 85\%.

\begin{figure*}
\begin{center}
\centerline{\includegraphics*[width=6.5in]{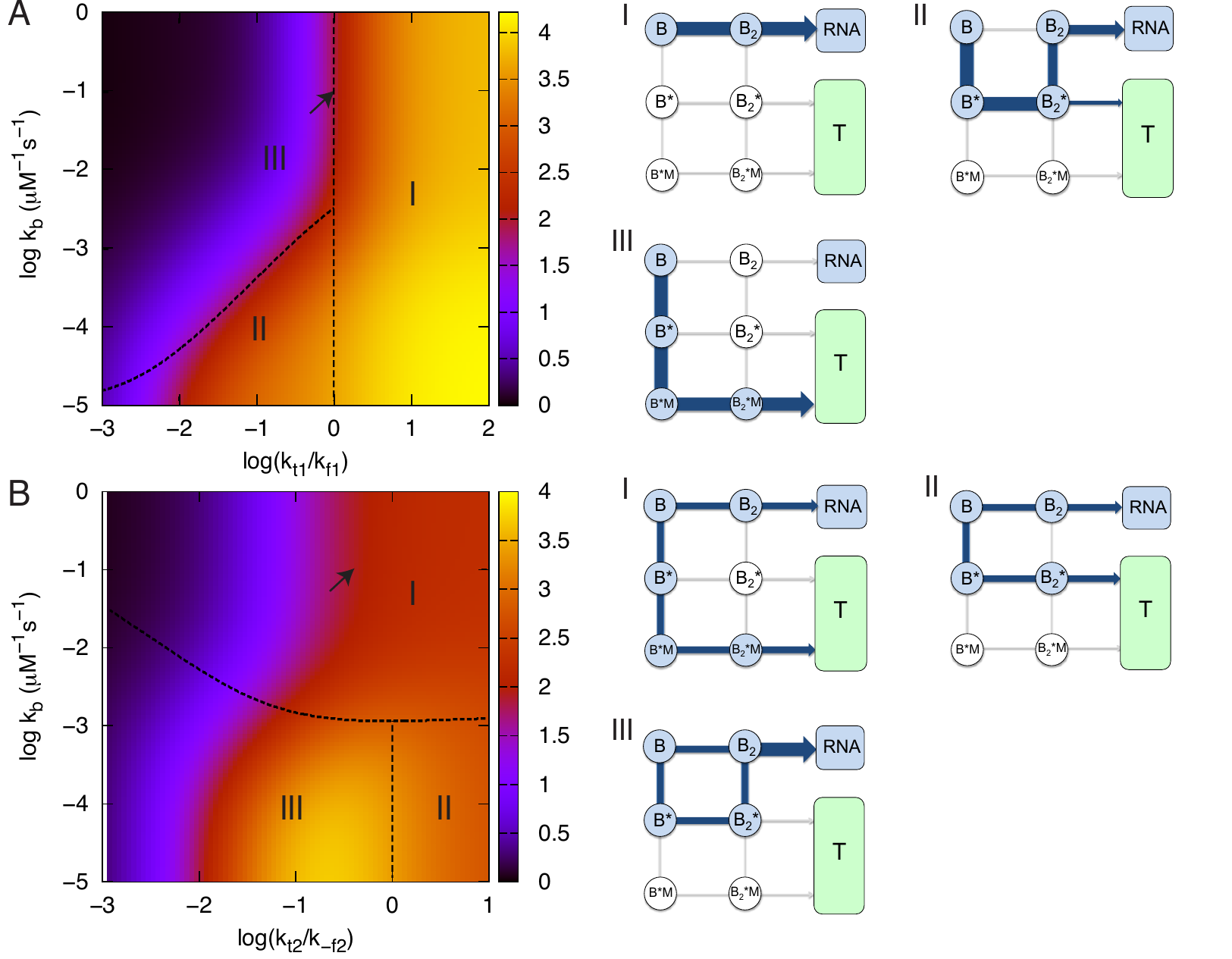}}
\caption{
Dependence of protein production on the network parameters with feedback.
(A) Protein levels as functions of $k_{t1}/k_{f1}$ and $k_{b}$ with
negative feedback using parameters in Tables I and II. The scale for the
$[P]$ production is shown in the color spectrum. The dependence of $[P]$ on
$k_{t1}$ and $k_{b}$ is catagorized into three 
regimes (see text for details). Points on the dashed line separating regime II
and regime III satisfy $k_{b}[M] = k_{t1}$.
The major pathway in thetranscription process in each regime is shown on the right.
The arrow indicates the data point
resulting from using the value of $k_{t1}$ and $k_{b}$ in Table I.
(B) Expression level of proteins as functions of $k_{t2}/k_{-f2}$ and
$k_{b}$ with negative feedback using the parameters in
Tables I and II. The dependence of $[P]$ 
on $k_{t2}$ and $k_{b}[M]$ is catagorized into three regimes. Points on the
dashed line separating regime I and regime II/III satisfy $k_{b}[M] = k_{-f2}$.
The corresponding major transcription pathways are shown on the right.
The data point corresponding to the arrow results from
using the value of $k_{t2}$ and $k_{b}$ in Table I. 
}
\label{Fig7} 
\end{center}
\end{figure*}

\subsection*{Dynamic phase diagram: Competition between folding, transcription, and binding rates}

To have a more complete picture on how the interplay between
folding of RNA transcripts, transcription, and metabolite binding regulate
the expression of $P$, we study the dependence of
$[P]$ on the transcription rates and the effective binding rate $k_{b}[M]$.
The dynamic phase diagram in Fig. 7 is calculated by varying both $k_{t1}$ 
($k_{t2}$) and $k_{b}$ with $K_{D}=10$ nM. 
In the first stage of transcription elongation, {\it i.e.} after the aptamer
sequence is transcribed, the formation of the aptamer structure is the key 
step in regulating transcription termination. Thus, the folding rate $k_{f1}$
and the effective metabolite binding rate are the key rates in competition with
$k_{t1}$ for regulation of $[P]$. Figure 7A shows three regimes for the
dependence of $[P]$ on $k_{t1}$ and $k_{b}[M]$. In regime I, $k_{t1} > k_{f1}$,
the folding rate is slow relative to transcription to the next stage (Fig. 1). 
Thus, the aptamer structure does not have enough time to form. The dominant 
flux  is from
$B$ to $B_{2}$, which leads to high probability of fully transcribed RNA
downstream because of the low transition rate from $B_{2}$ to $B_{2}^{*}$.
The metabolite binding has little effect on protein expression in this regime, 
particularly for large $k_{t1}/k_{f1}$, and hence the protein is highly 
expressed. In regime II, $k_{b}[M] < k_{t1} < k_{f1}$, the aptamer has enough 
time to fold but the metabolite binding is slow. The dominant flux is 
$B\rightarrow B^{*}\rightarrow B_{2}^{*}$, leading to formation of 
antitermination stem ($B_{2}^{*}\rightarrow B_{2}$) or transcription 
termination ($B_{2}^{*}\rightarrow B_{2t}^{*}$). The expression level of 
protein is thus mainly determined by $k_{-f2}$ and $k_{t2}$, and the protein
production is partially suppressed in this regime. 
In regime III, $k_{t1} < k_{f1}$ and $k_{t1} < k_{b}[M]$, the aptamer has 
sufficient time to both fold and bind metabolite, the dominant pathway is 
$B\rightarrow B^{*} \rightarrow B^{*}M \rightarrow B_{2}^{*}M$, leading to
transcription termination. The protein production is highly suppressed in this
regime. The results using parameters from Table I and Table II ($k_{t1}=k_{f1}$
and $k_{b}[M]/k_{f1}\sim 25$) fall 
on the interface of regime I and regime III, as shown by the arrow in Fig. 7A.
The metabolite binding does not reach thermodynamic equilibrium due to low
dissociation constant. However, the effective binding rate is high because the 
steady state concentration of metabolites ($\sim 25$ $\mu$M) is in large 
excess over RNA transcripts. Thus, the riboswitch is kinetically driven under 
this condition even when feedback is included.

With $k_{t1}$ comparable to $k_{f1}$, at the second stage of transcription 
elongation the key step against transcription termination
is the formation of the antiterminator stem ($B_{2}^{*}\rightarrow B_{2}$).
Figure 7B also shows three regimes for the dependence of $[P]$ on $k_{t2}$ and
$k_{b}[M]$ relative to $k_{-f2}$, and the associated most probable pathways are
displayed on the right. In regime I, $k_{b}[M] > k_{-f2}$,
the effective binding rate competes favorably with $B_{2}^{*}\rightarrow B_{2}$
transition so that $B_{2}$, if populated, is not likely to form 
the antiterminator stem. 
However, in this regime, the effective binding rate is also likely to
be larger than $k_{t1}$, resulting in most of the metabolite binding 
occurring at the first stage of transcription. Protein production is 
partially suppressed with the flux towards transcription termination flowing
through $B^{*}M\rightarrow B_{2}^{*}M\rightarrow T$ (terminated transcript).
In regime II, 
$k_{b}[M] < k_{-f2} < k_{t2}$, both the metabolite binding and 
$B_{2}^{*}\rightarrow B_{2}$ transition are too slow to occur.
The protein production level is mainly determined by $k_{t1}$ and $k_{f1}$.
The major pathways are $B\rightarrow B_{2} \rightarrow RNA$ and
$B\rightarrow B^{*}\rightarrow B_{2}^{*}\rightarrow T$,
leading to partial protein suppression.
In regime III, $k_{b}[M] < k_{-f2}$ and $k_{t2} < k_{-f2}$, 
metabolite binding is too slow to occur, but the riboswitch has enough time to
form the antiterminator stem before the terminator sequence is transcribed.
The major flux from $B_{2}^{*}$ flows to $B_{2}$, leading to fully
transcribed RNA, and proteins are highly expressed.
We note that using the parameters from Table I and Table II 
($k_{t2}/k_{-f2} = 0.4$ and $k_{b}M/k_{-f2} \sim 60$), the result falls in
regime I with partial protein suppresion. Among the three regimes, regime I
has efficient negative feedback, 
while the slow metabolite binding in regime II 
and regime III makes results resemble to those without feedback.
The dynamic phase diagrams predicts results with limiting cases of various
parameters, whose values may be within range {\it in vivo} and most certainly
{\it in vitro}. 
 
\begin{figure*}
\begin{center}
\centerline{\includegraphics*[width=6.5in]{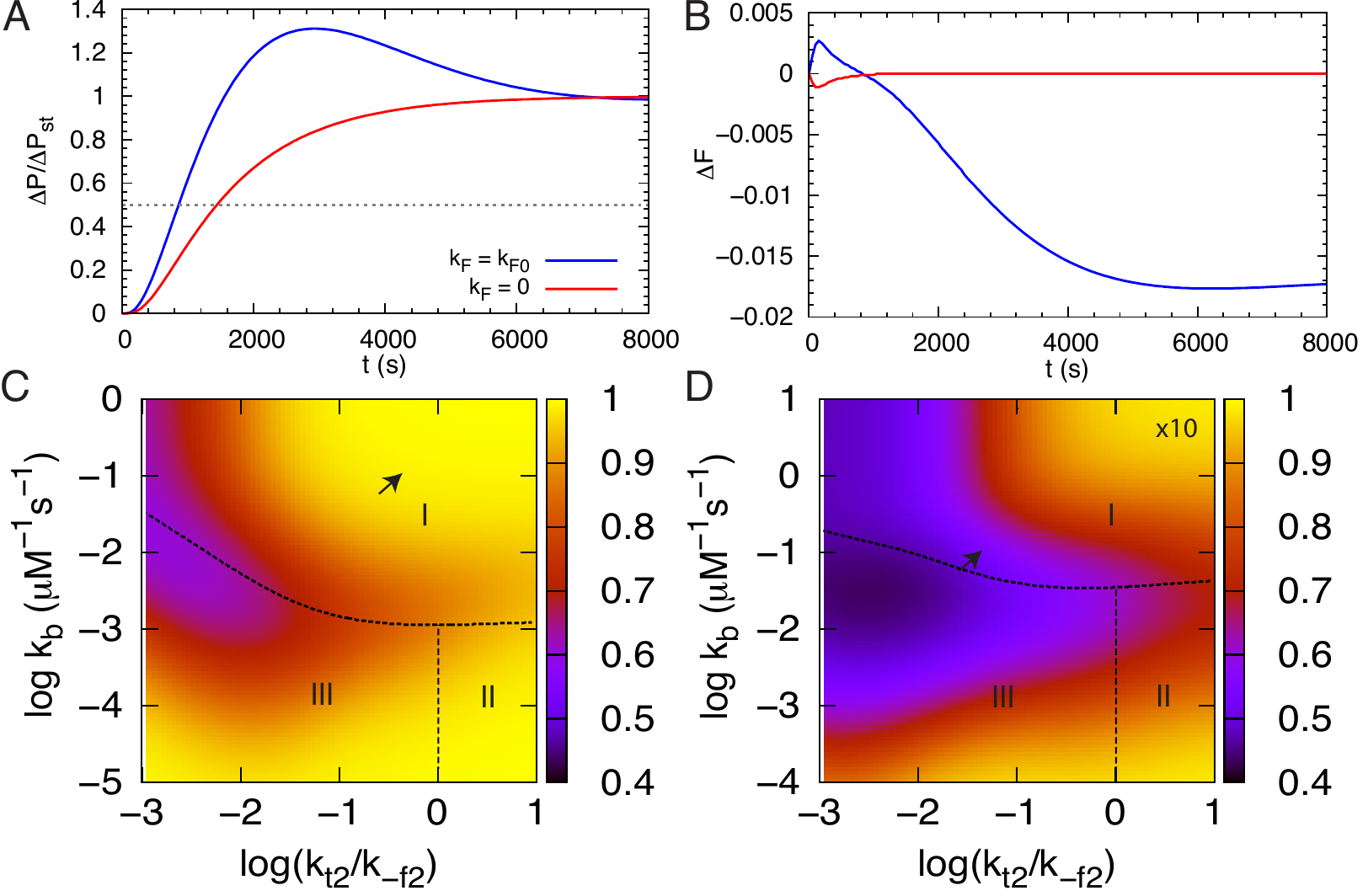}}
\caption{
Role of feedback to a spike in the DNA concentration.
(A) Response of protein level ($\Delta P$) relative to the change of
steady state level ($\Delta P_{st}$) when the DNA level increases
by $50\%$. The response time is 850 s with negative feedback (blue line)
and 1450 s without feedback (red line).
The values of folding and unfolding rates used are 10 fold larger than those
given in Table I.
(B) Response of the fraction of fully transcribed RNA,
$\Delta F$, when the DNA level increases by $50\%$.
The fraction of fully transcribed RNA increases initially with negative
feedback (blue line) and decreases initially without feedback (red line)
before settling to new steady state.
(C) Fractional change in protein steady state level relative
to that in DNA concentration, $\delta P_{st}/\delta D$, in repsonse to a $50\%$ rise in DNA level,as a function of $k_{t2}/k_{-f2}$ and $k_{b}$ for the riboswitch
network with negative feedback using paramters in Table I. Points on the dashed
line separating regime I and regime II/III satifsy $k_{b}[M] = k_{-f2}$.
(D) Same as (C) with the overall folding and unfolding rates being 10 fold
greater than those in Table I. The arrow indicates the data point
when $k_{t2}$ and $k_{b}$ are at the values from Table I.
}\label{Fig8}
\end{center}
\end{figure*}

\subsection*{Role of feedback in response to DNA bursts}
To assess how feedback affects the response to a sudden burst in DNA 
concentration we calculated the time dependent changes in the protein
concentration,
\begin{eqnarray}
\frac{\Delta P(t)}{\Delta P_{st}} = \frac{P_{[DNA]_{f}}(t)-P_{[DNA]_{i}}(t)}{P_{[DNA]_{f}}^{st}-P_{[DNA]_{i}}^{st}}
\end{eqnarray}
when the DNA concentration is switched from $[DNA]_{i}$ to $[DNA]_{f}$. 
In Eq. (4) $st$ stands for steady state.
Fig 8A shows the response time, defined as the time needed to 
reach half-way to the new steady-state level (dashed line in Fig. 8A), 
for $[DNA]_{f}$ = 3.75 nM and $[DNA]_{i}$ = 2.5 nM using folding and unfolding 
rates one order of magnitude larger than those in Table I. Small transient fluctuations in DNA concentration could arise from environmental stresses, and hence it is interesting to examine the response of the network to such changes. 
The values of the folding rates from Table I results in little difference in 
response time between cases with negative feedback and without feedback.
However, with larger folding rates, the response time for the systems with 
negative feedback is significantly shorter than without feedback (Fig. 8A). 
The fractional change of the fully transcribed RNA,
$\Delta F(t) \equiv f_{tra}([DNA]_{f},t)-f_{tra}([DNA]_{i},t)$ (Fig. 8B), 
shows a slight increase with overshoot initially before settling 
into a steady state level lower than the original one in the case of 
negative feedback (blue line in Fig. 8B).
For the case without feedback, the fraction of fully transcribed RNA 
decreases initially before reaching the expected steady state level 
(red line in Fig. 8B).

With negative feedback and 10-fold increase in overall 
folding and unfolding rates, the fractional increase in the protein steady 
state level, $\delta P_{st}\equiv (P_{[DNA]_{f}}^{st}-P_{[DNA]_{i}}^{st})/P_{[DNA]_{i}}^{st}$, 
in response to the increase in DNA level, 
$\delta D = ([DNA]_{f}-[DNA]_{i})/[DNA]_{i}$, is reduced by more than half of 
that in the case without feedback for certain range of parameters, as shown in 
Fig. 8C and 8D. Without feedback, $\delta P_{st} = \delta D$.
Negative feedback noticeably reduces the variations of expression in 
protein due to DNA level change. 
Substantial reduction occurs when the effective binding rate is 
comparable to $k_{-f2}$ and when $k_{t2} \leq k_{-f2}$(the interface between 
regime I and regime III in Fig. 8D). 

\section*{Discussion}

Transcription, regulated by metabolite binding to riboswitches, depends on an 
interplay of a number of time scales that are intrinsic to the 
co-transcriptional folding of the riboswitch as well as those determined by 
cellullar conditions. For a riboswitch to function  with a large dynamic range,
 transcription levels should change significantly as the metabolite concentration increases from a low to high value.
In the high concentration limit, RNA transcript  in 
the aptamer folded state binds a metabolite.  Low dissociation rate constant results in the formation of terminator stem, 
which subsequently terminates transcription.
In the low concentration limit, the aptamer folded state
is mostly unbound and can remain folded till transcription termination
or can fold to the antiterminator state leading to the transcription of the
full RNA. The levels of transcription termination is thus controlled by the 
transition rates between the aptamer folded and unfolded states. 

For {\it in vitro} description the efficiency of riboswitch is
determined by two conflicting requirements. If $\eta$, the dynamic range, is
to be maximized, then $\gamma_{2}$ has to be sufficiently low. However, at
low $\gamma_{2}$ and realistic values of the metabolite concentration,
$f_{ter}\approx 1$ which implies the switching function (needed to abort 
transcription) cannot be achieved. Thus, $\gamma_{2}$ has to have an optimal
range ($\gamma_{2}\sim (1-10)$) for the riboswitch to have sufficient dynamic
range without compromising the ability to switch from an on to off state.

In the presence of negative feedback loop the concentration of target metabolites is also regulated by gene 
expression. Under nominal operating conditions ($k_{t2}/k_{-f2}\sim 0.01-0.1$) binding of 
target metabolites, products of the downstream gene that 
riboswitches regulate, significantly suppresses the expression of proteins.
Negative feedback suppresses the protein level 
by about half relative to the case without feedback. 
{\it In vivo}, the presence of RNA binding proteins, such as NusA, may 
increase the pausing times, thus effectively reducing
the transcription rates. Thus, the repression of protein level by the 
riboswitch through metabolite binding may be up to 10 fold. Faster RNA folding and
unfolding rates than those we obtained may also increase the suppression 
by negative feedback and broaden the range of transcription rates over which
maximal suppression occurs. These predictions are amenable to experimental test.

In response to changes in the active operon level, the negative feedback  speeds up the response time of expression and modestly 
reduces  the percentage change in the protein level relative to 
change in operon level.  The steady state level of 
expression for autoregulation varies as a square-root of the DNA 
concentration.
Adaptive biological systems may minimize the variation in gene 
expression to keep the systems functioning normally even when the environments 
change drastically.  One may need to consider more complex networks than the single
autoregulation in the transcription network to find near perfect adaptation
to the environmental change \cite{Ma2009}.

Riboswitches provide novel ways to engineer biological 
circuits to control gene expression by binding small molecules. As found in
tandem riboswitches \cite{Sudarsan2006,Breaker2008}, 
multiple riboswitches can be engineered to
control a single gene with greater regulatory complexity or increase in the 
dynamic range of gene control. Synthetic riboswitches have been
successfully used to control the chemotaxis of bacteria \cite{Topp2007}. 
Our study provides a physical basis for not only analyzing
future experiments but also in anticipating their outcomes.

\section*{Acknowledgements}
We thank Michael Hinczewski for constructive suggestions and advice. 
This work was supported in part by a grant from the National Science Foundation
through grant number CHE09-10433.

%\section*{References}
% The bibtex filename
\bibliography{ref}

\end{document}